\documentclass[12pt,twoside]{article}

\usepackage{headerfo}
\usepackage{amsfonts}

\newcounter{theorem}
\newcommand{\newsection}[1]{{\setcounter{theorem}{0} \section{#1}}}
\newtheorem{Theorem}{Theorem}[section]
\newtheorem{Definition}[Theorem]{Definition}
\newtheorem{Proposition}[Theorem]{Proposition}
\newtheorem{Lemma}[Theorem]{Lemma}

\def\eop{{ \vrule height7pt width7pt depth0pt}\par\bigskip}

\newcommand{\R}{\mathbb R}
\newcommand{\C}{\mathbb C}

\newif\ifpdf
\ifx\pdfoutput\undefined
  \pdffalse
\else
  \pdfoutput=1
  \pdftrue
\fi

\usepackage{latexsym}

\chardef\aa=64

\begin{document}

\renewcommand{\author}{T.~Bloom \\  
Department of Mathematics \\
University of Toronto\\
Toronto, Ontario\\
Canada M5S 2E4,\\
\medskip L.~Bos \\  
Department of Computer Science\\
University of Verona \\
Verona, Italy\\
\medskip
and \\
N.~Levenberg\\
Department of Mathematics \\
Indiana University \\
Bloomington, Indiana, \\
USA\\}

\renewcommand{\title}{The Asymptotics of Optimal Designs for Polynomial Regression}
\newcommand{\stitle}{Optimal Designs}

\renewcommand{\date}{\today}
%%%%%%%%%%%%%%%%%%%%%%%%%%%%%%%%%%%%%%%%%%%%%%%%%%%%%%%%%%%%%%%%%%%%%%%%%%%%%%%
\flushbottom
\setcounter{page}{1}
\pageheaderlinetrue
\oddpageheader{}{\stitle}{\thepage}
\evenpageheader{\thepage}{\stitle}{}
\thispagestyle{empty}
\vskip1cm
\begin{center}
\LARGE{\bf \title}
\\[0.7cm]
\large{\author}
\end{center}
\vspace{0.3cm}
\begin{center}
\date
\end{center}
\begin{abstract}
We show that for a compact design space $K$ (which may in general be $K\subset\C^d$) the sequence of 
probability measures $\mu_s$
that give the so-called D--optimal (or, equivalently, G--optimal) experimental design for polynomial regression by multivariate
polynomials of degree $n,$ converges
weak--* to the equilibrium measure $\mu_K$ of Pluripotential Theory for $K.$
\end{abstract}

\setcounter{section}{0}

%%%%%%%%%%%%%%%%%%%%%%%%%%%%%%%%%%%%%%%%%%%%%%%%%%%%%%%%%%%%%%%%%%%%%%%%%%%%%%
%%%%%%%%%%%%%%%%%%%%%%%%%%%%%%%%%%%%%%%%%%%%%%%%%%%%%%%%%%%%%%%%%%%%%%%%%%%%%%
\newsection{Introduction} 

Optimal Experimental Design has a rich history within Statistics. The interested reader may consult the classical book of
Karlin and Studden [KS] (especially Chapter X), the more recent monograph of Dette and Studden [DS], or even the Wikipedia article [W], and the
references therein. As is well known (cf. [KS, Chap. X]), these optimal designs are equivalent to certain kinds of optimal
points for polynomial interpolation. Specifically,  D--optimal designs are equivalent to the so-called Fekete points of the set $K.$ What
may have received less attention in the Statistics literature is the fact that the asymptotics of such point systems (measures) is in
one variable intimately connected to classical Complex Potential
Theory and in several variables, as it turns out, to Complex Pluripotential Theory. 

In recent years much progress has been made in this latter field and, in particular, it allows the determination of the
limit of the sequence of D--optimal measures as the degree of the polynomials on which regression takes place goes to
infinity. The purpose of this note is to explain these developments and how they apply to Statistics. We begin with a review of the optimal designs that we wish to discuss.

In practical applications the domains of interest are typically real cubes, balls, simplices or other nice geometric sets. 
But the theory is very general, and hence
we take for our design space a compact $K\subset \C^d$ (later we will impose a mild regularity condition on $K$). We let ${\cal P}_s(K),$ denote the set of polynomials of degree $s$ restricted to $K$ 
and set $n:={\rm dim}({\cal P}_s(K)).$

We may write any $p\in {\cal P}_s(K)$ in the form
\[p=\sum_{k=1}^n \theta_k p_k\]
where ${\cal B}_s:=\{p_1,p_2,\ldots,p_n\}$ is a basis for ${\cal P}_s(K).$

Suppose now that we observe the values of a particular
$p\in {\cal P}_s(K)$ at a set of  $m\ge n$ points $X:=\{x_j\,:\, 1\le j\le m\}\subset K$ with some random errors, i.e., we observe
\[y_j =p(x_j)+\epsilon_j,\quad 1\le j\le m\]
where we assume that the errors $\epsilon_j\sim N(0,\sigma)$ are independent. In matrix form this becomes
\[ y= V_s\theta+\epsilon\]
where $y,\theta,\epsilon\in\C^N$ and
\[V_s=\left[\begin{array}{cccccc}
p_1(x_1)&p_2(x_1)&\cdot&\cdot&\cdot&p_n(x_1) \cr
p_1(x_2)&p_2(x_2)&\cdot&\cdot&\cdot&p_n(x_2) \cr
\cdot&&&&&\cdot\cr
\cdot&&&&&\cdot\cr
\cdot&&&&&\cdot\cr
\cdot&&&&&\cdot\cr
\cdot&&&&&\cdot\cr
p_1(x_m)&p_2(x_m)&\cdot&\cdot&\cdot&p_n(x_m) \end{array}\right]\in \C^{m\times n}\]
is the associated Vandermonde matrix.

Our assumption on the error vector $\epsilon$ means that
\[{\rm cov}(\epsilon)=\sigma^2I_n\in\R^{n\times n}.\]
Now, the least squares estimate of $\theta$ is
\[\widehat{\theta}:=(V_s^*V_s)^{-1}V_s^*y\]
and we may compute the covariance matrix 
\[{\rm cov}(\widehat{\theta})=\sigma^2(V_s^*V_s)^{-1}.\]
Hence the confidence region of level $t$ for $\theta$ is the set
\begin{eqnarray*}
&&\{\theta\in \C^n\,:\,(\theta-\widehat{\theta})^*[{\rm cov}(\widehat{\theta})]^{-1}(\theta-\widehat{\theta})\le t\} \\
&=&\{\theta\in \C^n\,:\,\sigma^{-2}(\theta-\widehat{\theta})^*(V_s^*V_s)(\theta-\widehat{\theta})\le t\}.
\end{eqnarray*}
The volume of such a set is proportional to $1/\sqrt{{\rm det}(V_s^*V_s)}$ and hence maximizing the
${\rm det}(V_s^*V_s)$ is equivalent to choosing the observation points $x_i\in K$ so as to have the most
``concentrated'' confidence region for the parameter to be estimated.

Note however that the entries of $\displaystyle{ {1\over m}V_s^*V_s}$ are the discrete inner products of the $p_i$
with respect to the measure
\begin{equation}\label{StatsMeas}
\mu_X ={1\over m}\sum_{k=1}^m \delta_{x_k}.
\end{equation}
More specifically,
\[ {1\over m}V_s^*V_s=M_s(\mu_X)\]
where
\begin{equation}\label{M}
M_s(\mu):=\left[\int_K\overline{p_i(z)}p_j(z)d\mu\right]\in\C^{n\times n}
\end{equation}
is the  Moment, or Gram, matrix of the polynomials $p_i$ with respect to the measure $\mu.$

In general we may consider arbitrary probability measures on $K,$ setting
\[{\cal M}(K):=\{\mu\,:\, \mu \,\,\hbox{is a probability measure on}\,\, K\}.\]

\begin{Definition}\label{Doptimal}
A probability measure (or design) $\mu\in {\cal M}(K)$ is said to be a D--optimal measure of degree $s$ if it has the
property that
\[{\rm det}(M_s(\mu))\ge {\rm det}(M_s(\xi)),\,\,\forall \xi\in{\cal M}(K).\]
\end{Definition}

There is also a second statistical interpretation of D--optimal measures.
If we set
\begin{equation}\label{P}
{\mathbf p}(z)=\left[\begin{array}{c}p_1(z)\cr p_2(z)\cr\cdot\cr\cdot\cr p_n(z)\end{array}\right]\in \C^n
\end{equation}
then the least squares estimate of the observed polynomial is
\[{\mathbf p}^t(z)\widehat{\theta}.\]
We may compute its variance to be
\begin{eqnarray}
{\rm var}({\mathbf p}^t(x)\widehat{\theta})&=&\sigma^2{\mathbf p}^*(z)(V_s^*V_s)^{-1}{\mathbf p}(z) \nonumber\\
&=&{1\over m}\sigma^2 {\mathbf p}^*(z)(M_s(\mu_X))^{-1}{\mathbf p}(z)\label{variance}
\end{eqnarray}
where $\mu_X$ is again given by (\ref{StatsMeas}). 

In the Statistics literature (see e.g. [DS]) one usually denotes, for $\mu\in{\cal M}(K),$ 
\[G_s(\mu)=\max_{z\in K}{\mathbf p}^*(z)(M_s(\mu))^{-1}{\mathbf p}(z).\]

\begin{Definition}\label{Goptimal}
A probability measure $\mu\in {\cal M}(K)$ is said to be a G--optimal measure of degree $n$ if it has the
property that
\[ G_s(\mu)\le G_s(\xi),\,\,\,\forall \xi\in{\cal M}(K) .\]
\end{Definition}

It follows from (\ref{variance}) that a G--optimal measure minimizes the maximum variance of
the estimate of the observed polynomial. 

The remarkable Kiefer-Wolfowitz equivalence theorem states that these two notions of optimality are
equivalent.

\begin{Theorem}(Kiefer and Wolfowitz [KW]) 
A measure $\mu\in{\cal M}(K)$ with ${\rm det}(M_s(\mu))\neq 0$ is G--optimal of degree $s$ if and only if it
is D--optimal of degree $s.$
\end{Theorem}

The G-optimality criterion has also an interpretation in terms of the polynomials orthogonal on $K$
with respect to the measure $\mu.$ To see this, suppose that $M_s(\mu)$ is non-singular and note that 
then the matrix $M_s(\mu),$ being a Gram matrix, is
positive definite.  It's inverse is then also positive definite and 
hence has a Cholesky factorization $(M_s(\mu))^{-1}=L_s(\mu)^*L_s(\mu)$ where $L_s(\mu)\in \C^{n\times n}$
is lower triangular. It follows that we may write
\begin{eqnarray*}
{\mathbf p}^*(z)(M_s(\mu))^{-1}{\mathbf p}(z)&=&{\mathbf p}^*(z)L_s(\mu)^*L_s(\mu){\mathbf p}(z) \\
&=&(L_s(\mu){\mathbf p}(z))^*(L_s(\mu){\mathbf p}(z)) \\
&=&\sum_{j=1}^n|q_j(z)|^2
\end{eqnarray*}
where 
\[{\mathbf q}:=\left[\begin{array}{c} q_1 \cr q_2\cr \cdot \cr \cdot \cr q_n\end{array}\right]:=L_s(\mu)
\left[\begin{array}{c} p_1 \cr p_2\cr \cdot \cr \cdot \cr p_n\end{array}\right].\]
The polynomials $q_j$ are in fact orthonormal as
\begin{eqnarray*}
\left[\int_K q_i(z)\overline{q_j(z)}d\mu(z)\right]&=&\int_K {\mathbf q}(z){\mathbf q}(z)^* d\mu(z) \\
&=&\int_K (L_s(\mu){\mathbf p}(z))(L_s(\mu){\mathbf p}(z))^* d\mu(z) \\
&=&\int_K L_s(\mu){\mathbf p}(z){\mathbf p}(z)^* L_s(\mu)^*  d\mu(z) \\
&=&L_s(\mu)\left(\int_K {\mathbf p}(z){\mathbf p}(z)^* d\mu(z) \right) L_s(\mu)^*  \\
&=&(L_s(\mu)M_s(\mu)L_s(\mu)^*)^* \\
&=&I.
\end{eqnarray*}
Indeed, since $L_s(\mu)$ is lower triangular, the $q_j$ are the just the result of 
applying the Gram-Schmidt orthonormalization procedure to the $p_j.$

Now note that
\begin{equation}\label{Kn}
K_s^\mu(z):=\sum_{j=1}^n|q_j(z)|^2
\end{equation}
is the diagonal of the reproducing kernel for ${\cal P}_s(K)$ (with respect to the measure $\mu$) 
and is sometimes also called the
(reciprocal of the) Christoffel function. It plays an important role in the theory of Orthogonal Polynomials.

Hence
\begin{equation}\label{SigOfKn}
{\rm var}({\mathbf p}^t(z)\widehat{\theta})={1\over m}\sigma^2 K_s^\mu(z)
\end{equation}
and the experiment that minimizes the maximum variance of the estimate of the observed polynomial is exactly the
one that minimizes the maximum of $K_s^\mu.$

For each $s$ there is a D--optimal measure $\mu_s.$ We will show that these $\mu_s$ converge (in the weak--* sense) to
what is called the equilibrium measure of Complex Pluripotential Theory for $K.$ In order to make this more precise we
will need first of all to introduce some of the basic notions of this theory.
We refer the reader to the monograph [K] and also to Appendix B of [ST] for more details.

\begin{Definition}
A function $u\,:\,\C^d\to[-\infty,\infty)$ is said to be plurisubharmonic (psh) if it is upper semi-continuous (usc)
and, when restricted to any complex line, is either subharmonic or identically $-\infty.$
\end{Definition}

\begin{Definition}
A set $K\subset \C^d$ is said to be \emph{pluripolar} if there exists a psh function $u,$ not identically $-\infty,$ 
such that
$u(z)=-\infty$ for all $z\in K.$ 
\end{Definition}

(Pluri)polar sets are in some sense the inconsequential sets of Potential Theory and are too ``small'' for there
to be a reasonable theory. A prototypical example of a psh function is $u=\log|f|$ where $f$ is holomorphic on 
$\C^d.$ In particular, for such a $u,$ $u\equiv -\infty$ on the zero set of $f,$ which is therefore a
pluripolar set. More generally, any complex analytic variety (other than all of $\C^d$) is pluripolar.
We will henceforth make the assumption that the design space $K$ is \emph{non-pluripolar}.

In Section 3 below we will state the convergence theorem. But first we will show that the notions of D--optimal and
G--optimal can be generalized to the so-called weighted case. The introduction of such a weight is a crucial step in the
proof of the main theorem. We note that these results have appeared elsewhere (see [BBLW]) but here we offer much simplified
proofs, based on the integral formulas of Lemma \ref{DetasInt}, which are adapted from standard results used in Random Matrix Theory (see
e.g. [D]).

\newsection{Weighted Optimal Designs}

Consider a design space 
$K\subset \C^d,$ compact and non-pluripolar. 

\begin{Definition}
A function $w\,:\,K\to \R$ is said to be an admissible weight function if\par
\noindent (i) $w\ge 0$ on $K$\par
\noindent (ii) $w$ is upper semi-continuous\par
\noindent (iii) the set
\[\{z\in K\,:\, w(z)>0\}\]
is not pluripolar.
\end{Definition}
For $\mu$ a probability measure on $K$ and admissible weight $w$ we denote the associated weighted inner product
of degree $s$ by
\begin{equation}\label{Wip}
\langle f,g\rangle_{\mu,w}:=\int_K \overline{f(z)}g(z)w^{2s}(z)d\mu.
\end{equation}
For a (fixed) basis ${\cal B}_s=\{p_1,p_2,\cdots, p_n\}$ of ${\cal P}_s(K)$ we form the weighted Moment (Gram) matrix
\begin{equation}\label{WGramMatrix}
M_s^{\mu,w}=M_s^{\mu,w}({\cal B}_s):=[\langle p_i,p_j\rangle_{\mu,w}]\in\C^{n\times n}
\end{equation}
and the associated weighted Christoffel function
\begin{equation}\label{WKn}
K_s^{\mu,w}(z):=\sum_{j=1}^n|q_j(z)|^2w^{2s}(z)=({\mathbf p}(z))^*(M_s^{\mu,w})^{-1}{\mathbf p}(z)
\end{equation}
where, as before, $Q_n=\{q_1,q_2,\cdots,q_n\}$ is an orthonormal basis for ${\cal P}_s(K)$ with respect
to the inner-product (\ref{Wip}). We note that as the Christoffel function is (essentially) the diagonal of the reproducing
kernel, it is independent of the particular orthonormal basis $Q_s.$ For its statistical significance see (\ref{SigOfKn}).

\begin{Definition}\label{WeigthedDoptimal} Suppose that $w$ is an admissible weight on $K.$
A probability measure (or design) $\mu\in {\cal M}(K)$ is said to be a weighted D--optimal measure of degree $s$ if it has the
property that
\[{\rm det}(M_s^{\mu,w})\ge {\rm det}(M_s^{\xi,w}),\,\,\forall \xi\in{\cal M}(K).\]
\end{Definition}

\begin{Definition}\label{WeightedGoptimal} Suppose that $w$ is an admissible weight on $K.$
A probability measure $\mu\in {\cal M}(K)$ is said to be a weighted G--optimal measure of degree $s$ if it has the
property that
\[ \max_{z\in K} K_s^{\mu,w}(z) \le \max_{z\in K} K_s^{\xi,w}(z),\,\,\,\forall \xi\in{\cal M}(K) .\]
\end{Definition}

By (the proof of) Lemma 2.1 of [KS], Chapter X], the set of matrices
\[\{M_s^{\mu,w}\,:\,\mu\,\,\hbox{is a probability measure on}\,\,K\}\]
is compact (and convex). Hence  D--optimal and G--optimal measures of
degree $s$ for $K$ and $w$ always exists. They will not in general be unique. 

We recall that for a basis ${\cal B}_s$ and a set of points $X_s=\{z_i\,:\,1\le i\le n\}\subset K$ the matrix
\begin{equation} \label{vdmMatrix}
V_s(z_1,z_2,\cdots,z_n)=V_s({\cal B}_s,X_s)=[p_j(z_i)]\in \C^{n\times n}
\end{equation}
is called the Vandermonde matrix of the system.  We will let
\begin{equation} \label{vdmDet}
VDM(z_1,z_2,\cdots,z_n)=VDM_{{\cal B}_s}(z_1,\cdots,z_n):={\rm det}(V_s(z_1,\cdots,z_n))
\end{equation}
denote its determinant.

We will make use of the following two formulas that express 
${\rm det}(M_s^{\mu,w})$ and $K_s^{\mu,w}$ in terms of these 
Vandermonde determinants.

\begin{Lemma}\label{DetasInt}
Suppose that $\mu\in{\cal M}(K)$ and that $w$ is an admissible weight. Then
 (cf. formula (3.3) of [BL])
\begin{eqnarray}
&{\rm det}(M_s^{\mu,w})= &\label{DetasInt1} \\
&\displaystyle{\frac{1}{n!}\int_{K^n}\!\!|VDM(z_1,\cdots,z_n)|^2w(z_1)^{2s}\cdots w(z_n)^{2s} d\mu(z_1)\cdots d\mu(z_n).}& \nonumber
\end{eqnarray}
and
\begin{eqnarray}
&K_s^{\mu,w}(z)=& \label{DetasInt2}\\
&\displaystyle{\frac{n}{Z_n}\int_{K^{n-1}}\!\!
|VDM(z,z_2,\cdots,z_n)|^2w(z)^{2s}w(z_2)^{2s}\cdots w(z_n)^{2s}d\mu(z_2)\cdots d\mu(z_n)}& \nonumber
\end{eqnarray}
where 
\begin{eqnarray*}
Z_n&:=&n!\,{\rm det}(M_s^{\mu,w})\\
&=&\int_{K^n}|VDM(z_1,\cdots,z_n)|^2w(z_1)^{2s}\cdots w(z_n)^{2s} d\mu_s(z_1)\cdots d\mu_s(z_n).
\end{eqnarray*}
\end{Lemma}
\noindent{\bf Proof.} 
As before, let $L_s\in\C^{n\times n}$ be a lower triangular Choleski factor of $(M_s^{\mu_s,w})^{-1},$ i.e., such that
\[(M_s^{\mu_s,w})^{-1}=L_s^*L_s.\]
It follows then that the basis $Q_s:=\{q_1,q_2,\cdots,q_n\}$ given by
\[ \left[\begin{array}{c} q_1 \cr q_2\cr \cdot \cr \cdot \cr q_n\end{array}\right]:=L_s(\mu)
\left[\begin{array}{c} p_1 \cr p_2\cr \cdot \cr \cdot \cr p_n\end{array}\right] \]
is orthonormal with respect to the inner product (\ref{Wip}).

It is elementary to verify the basis transition formula for Vandermonde matrices,
\[V_s({\cal B}_s)=V_s(Q_s)L_s^{-t}\]
and hence that
\begin{eqnarray*}
|{\rm det}(V_s({\cal B}_s))|^2&=&|{\rm det}(V_s({\cal B}_s))|^2{\rm det}(L_s^{-1}(L_s^{-1})^*)\\
&=&|{\rm det}(V_s({\cal B}_s))|^2{\rm det}(M_s^{\mu,w}).
\end{eqnarray*}
In other words,
\begin{eqnarray}
 |VDM(z_1,\cdots,z_n)|^2&=&|VDM_{{\cal B}_s}(z_1,\cdots,z_n)|^2 \nonumber\\
 &=&{\rm det}(M_s^{\mu,w})|VDM_{Q_s}(z_1,\cdots,z_n)|^2. \label{newBasis}
\end{eqnarray}
Now, by the Leibniz formula for determinants,
\begin{equation}
VDM_{Q_s}(z_1,\cdots,z_n)=\sum_\sigma {\rm sgn}(\sigma)\prod_{i=1}^nq_{\sigma(i)}(z_i),\label{leibniz}
\end{equation}
and by orthogonality,
\begin{eqnarray*}
&&\int_{K^n}\!\!|VDM_{Q_s}(z_1,\cdots,z_n)|^2w(z_1)^{2s}\cdots w(z_n)^{2s} d\mu(z_1)\cdots d\mu(z_n)\\
&=&\int_{K^n}\!\!\overline{VDM_{Q_s}(z_1,\cdots,z_n)}VDM_{Q_s}(z_1,\cdots,z_n)w(z_1)^{2s}\cdots w(z_n)^{2s} d\mu(z_1)\cdots d\mu(z_n)\\
&=&n!
\end{eqnarray*}
as there are $n!$ permutations. The formula (\ref{DetasInt1}) now follows from this and (\ref{newBasis}).

The proof of (\ref{DetasInt2}) is very similar. Using again the Leibniz formula (\ref{leibniz}), we easily see that
\begin{eqnarray*}
&\displaystyle{\int_{K^{n-1}}\!\! |VDM(z,z_2,\cdots,z_n)|^2w(z)^{2s}w(z_2)^{2s}\cdots w(z_n)^{2s}d\mu(z_2)\cdots d\mu(z_n)}&\\
&\displaystyle =\sum_{j=1}^n C_j|q_j(z)|^2w(z)^{2s}&
\end{eqnarray*}
for some constants $C_j.$ But by symmetry, all the $C_j$ must be the same, i.e.,
\begin{eqnarray*}
&\displaystyle \int_{K^{n-1}}\!\!
|VDM(z,z_2,\cdots,z_n)|^2w(z)^{2s}w(z_2)^{2s}\cdots w(z_n)^{2s}d\mu(z_2)\cdots d\mu(z_n)&\\
&\displaystyle =CK_s^{\mu,w}(z)&
\end{eqnarray*}
for some constant $C.$ The value of the normalization constant $C$ given by (\ref{DetasInt2}) follows by 
integrating both sides over $K$ with respect to $w(z)^{2s}d\mu(z)$ and noting that
\[\int_K  K_s^{\mu,w}(z) w(z)^{2s}d\mu(z)=\int_K \sum_{j=1}^n |q_j(z)|^2w(z)^{2s}d\mu(z)=n\]
by the orthonormality of the $q_j.$
\eop

The Kiefer-Wolfowitz equivalence theorem also holds in the weighted case; 
a proof is given in Prop. 3.1 of [BBLW]. Here we offer a somewhat simplified proof
based on the integral formuals of Lemma \ref{DetasInt}.

\begin{Proposition} (Weighted Kiefer--Wolfowitz [KW])
\label{KW}
Let $w$ be an admissible weight on $K.$ A probability measure $\mu$
for which $M_s^{\mu,w}$ is non-singular
is a weighted D--optimal measure of degree $s$ 
if and only if it is a weighted G--optimal measure of degree $s$ 
with the property that
\[\displaystyle{  \max_{z\in K}K_s^{\mu,w}(z)=n}.\]
\end{Proposition} 
\noindent {\bf Proof.} First note, as already observed by Kiefer and Wolfowitz, 
the functional
\[\mu\mapsto \log({\rm det}(M_s^{\mu,w}))\]
is concave. Indeed this follows easily from
the fact that, for $\mu_t:=t\mu_1+(1-t)\mu_0,$
\[M_s^{\mu_t,w}=tM_s^{\mu_1,w}+(1-t)M_s^{\mu_0,w};\]
see either [KW] or else [Bo] for the details.

Hence, a probabilty measure $\mu_0$ is D-optimal iff $h'(0)\le0$ for
\[h(t)=h(t;\mu_0,\mu_1):=\log({\rm det}(M_s^{\mu_t,w}))\]
and all $\mu_1\in {\cal M}(K).$

But, from the formula (\ref{DetasInt1}), $h(t)$ is the logarithm of a polynomial of degree $n$ in $t$ and
one may easily compute 
\begin{equation}\label{hprime}
h'(0)=\int_K K_s^{\mu_0,w}(z)d\mu_1(z)-n.
\end{equation}

Assume now that $\mu_0$ is D--optimal. For $z\in K$ take $\mu_1=\delta_z,$ the measure supported 
at $z.$ Then (\ref{hprime}) becomes
\[K_s^{\mu_0,w}(z)\le n\]
and since $z\in K$ was arbitrary,
\begin{equation} \label{Maxformu0}
\max_{z\in K}K_s^{\mu_0,w}\le n.
\end{equation}
Since 
\[ \int_K K_s^{\xi,w} w(z)^{2s}d\xi(s)=\int_K \sum_{j=1}^n |q_j(z)|^2w(z)^{2s} d\xi(z)=n\]
we must  have $\displaystyle \max_{z\in K}K_s^{\xi,w}(z)\ge n$
for all $\xi\in {\cal M}(K).$ In other words, by (\ref{Maxformu0}), $\mu_0$ is G-optimal (cf. Definition \ref{WeightedGoptimal})
and 
\[\max_{z\in K}K_s^{\mu_0,w}(z)=n.\]

Conversely, suppose now that $\mu_0\in{\cal M}(K)$ is G--optimal with the property that
\[\max_{z\in K}K_s^{\mu_0,w}=n.\]
Then, from (\ref{hprime}) we have 
\[h'(0)=\int_K K_s^{\mu_0,w}(z) d\mu_1(z)-n\le \int_K n\,d\mu_1(z)-n=n-n=0,\]
for all $\mu_1\in {\cal M}(K),$
and we see that $\mu_0$ is also then D--optimal. \eop

\medskip From now on we will say that a measure is {\it optimal} if it is either D--optimal or, equivalently, G--optimal.

The last statement of the preceeding proposition yields the following key property of optimal measures.

\begin{Lemma}\label{MaxIsN}
Suppose that $\mu$ is optimal for $K$ and $w.$ Then
\[K_s^{\mu,w}(z)=n,\quad a.e.\,\, [\mu].\]
\end{Lemma}

\noindent {\bf Proof.} On the one hand
\[\max_{z\in K}K_s^{\mu,w}(z)=n\]
while on the other hand, again by orthonormality of the $q_j,$
\[ \int_K K_s^{\mu,w}\, d\mu=\int_K\sum_{j=1}^n|q_j(z)|^2w(z)^{2s}\,d\mu(z)=n,\]
and the result follows. \eop

\medskip
Of fundamental importance for us will be
\begin{Definition}
Suppose that $K\subset \C^d$ is compact and that $w$ is an admissible weight function 
on $K.$ We set
\[\delta_s^w(K):=\left(\max_{z_i\in K}|VDM(z_1,\cdots,z_n)|w^s(z_1)w^s(z_2)\cdots w^s(z_N)\right)^{1/m_n}\]
where $m_s=dsn/(d+1)$ is the sum of the degrees of the $n$ monomials of degree at most $s.$
Then 
\[\delta^w(K)=\lim_{s\to\infty} \delta_s^w(K)\]
is called the weighted transfinite diameter of $K.$ We refer to $\delta_s^w(K)$ as the 
weighted  $s$th order diameter of $K.$
\end{Definition}
A proof that this limit exists may be found in [BL] or [BB1]; it was first proved in the unweighted case ($w\equiv 1$; i.e., $\delta^1(K)$) by Zaharjuta [Z].

Given the close connection between Vandermonde matrices and Gram matrices,
it is perhaps not suprising that we have
\begin{Proposition}\label{tfd}
Suppose that $K$ is compact and that $w$ is an admissible weight function. Suppose further that
$\mu_s$ is an optimal measure of degree $s$ for $K$ and $w.$ Take the basis ${\cal B}_s$ to be the standard
basis of monomials for ${\cal P}_s.$ Then
\[\lim_{s\to\infty}{\rm det}(M_s^{\mu_s,w})^{1/(2m_s)}=\delta^w(K).\]
\end{Proposition}
\noindent {\bf Proof.} This is Proposition 4.3 of [BBLW]. \eop 

\medskip
Of course, it then follows that
\begin{equation}\label{limgn}
\lim_{s\to \infty}{1\over 2 m_s}\log\,{\rm det}(M_s^{\mu_s,w})=\log(\delta^w(K)).
\end{equation}
Now, suppose that $u\in C(K)$ and that $w$ is an admissible weight function. Following the ideas in [Be, BB1, BB2, BN, BBN] we consider the weight $w_t(z):=w(z)\exp(-tu(z)),$ $t\in\R,$ and let $\mu_s$ be an optimal measure of degree $s$ for $K$ and $w.$
We set
\begin{equation}\label{fn}
f_s(t):=-{1\over 2 m_s}\log\,{\rm det}(M_s^{\mu_s,w_t}).\end{equation}
For $t=0,$ $w_0=w$ and (\ref{limgn}) says
\[\lim_{s\to\infty}f_s(0)=-\log(\delta^{w}(K)).\]

We have the following formula for the derivative of $f_s.$
\begin{Lemma}\label{1stderiv}We have
\[ f_s'(t)={d+1\over dn}\int_K u(z)K_s^{\mu_s,w_t}(z)d\mu_s.\]
In particular,
\begin{eqnarray}
f_s'(0)&=&{d+1\over dn}\int_K u(z)K_s^{\mu_s,w}(z)d\mu_s \nonumber \\
&=& {d+1\over d}\int_K u(z)d\mu_s \quad \hbox{(by Lemma \ref{MaxIsN})}.
\end{eqnarray}
\end{Lemma}
\noindent {\bf Proof.} This was proved by different means in 
[Be, Lemma 6.4] and [BBLW, Lemma 3.5]. The proof we offer here is
based on the integral formulas of Lemma \ref{DetasInt}.

By  (\ref{DetasInt1}) we may write
\[ f_s(t)=-{1\over 2 m_s}\log(F_s)-{1\over 2 m_s}\log(n!)\]
where
\[F_s(t):=\int_{K^n}V\exp(-tU)d\mu\]
and
\begin{eqnarray*}
V&:=&V(z_1,z_2,\cdots,z_n)=|VDM(z_1,\cdots,z_n)|^2w(z_1)^{2s}\cdots w(z_n)^{2s}, \\
U&:=&U(z_1,z_2,\cdots,z_n)=2s(u(z_1)+\cdots+u(z_n)), \\
d\mu&:=&d\mu_s(z_1)d\mu_s(z_2)\cdots d\mu_s(z_n).
\end{eqnarray*}
Further, by (\ref{DetasInt2}) for $w=w_t$ and $\mu=\mu_s,$ we have
\[ K_s^{\mu_s,w_t}(z)=\frac{n}{Z_n}\int_{K^{n-1}}
V(z,z_2,z_3,\cdots,z_n)\exp(-tU)d\mu_s(z_2)\cdots d\mu_s(z_n)\]
where 
\[Z_n=Z_n(t):=n!\,{\rm det}(M_s^{\mu_s,w_t})=\int_{K^n}V\,\exp(-tU) d\mu.\]
Note that $Z_n(t)=F_s(t).$
Now
\[f_s'(t)=-{1\over 2 m_s}\frac{F_s'(t)}{F_s(t)}\]
and we may compute
\begin{eqnarray*}
F_s'(t)&=&\int_{K^n} V(-U)\exp(-tU)d\mu_s(z_1)\cdots d\mu_s(z_n)\\
&=& -2s\int_{K^n} (u(z_1)+\cdots+u(z_n))V\exp(-tU)d\mu_s(z_1)\cdots d\mu_s(z_n).
\end{eqnarray*}
Notice that the integrand is symmetric in the variables and hence we
may ``de-symmetrize'' to obtain
\[F_s'(t)=-2sn\int_{K^n}u(z_1)V(z_1,\cdots,z_n)\exp(-tU)d\mu_s(z_1)\cdots d\mu_s(z_n)\]
so that, integrating in all but the $z_1$ variable, we obtain
\[F_s'(t)=-2sn\int_K u(z)K_s^{\mu_s,w_t}(z)\frac{Z_n}{n}d\mu_s(z).\]
Thus, using the fact that $Z_n(t)=F_s(t),$ we obtain
\begin{eqnarray*}
f_s'(t)&=&-{1\over 2 m_s}\frac{F_s'(t)}{F_s(t)}\\
&=&\frac{1}{2 m_s}(2s)\int_K u(z)K_s^{\mu_s,w_t}(z)d\mu_s(z)\\
&=&\frac{s}{dsn/(d+1)}\int_K u(z)K_s^{\mu_s,w_t}(z)d\mu_s(z) \\
&=&\frac{d+1}{dn}\int_K u(z)K_s^{\mu_s,w_t}(z)d\mu_s(z),
\end{eqnarray*}
as claimed.
In particular,
\begin{eqnarray*}
f_s'(0)&=&\frac{d+1}{dn}\int_K u(z)K_s^{\mu_s,w}(z)d\mu_s(z)\\
&=&\frac{d+1}{d}\int_K u(z)d\mu_s(z)
\end{eqnarray*}
by Lemma \ref{MaxIsN}. \eop

The next result was proved in a different way in [BBN, Lemma 2.2] and also
in [BBLW, Lemma 3.6].

\begin{Lemma}\label{2ndderiv}
The functions $f_s(t)$ are concave.  
\end{Lemma}
\noindent {\bf Proof.} Since $f_s(t)$ is twice differentiable we need only to show
that $f_s''(t)\le0.$ Now, with the notation used in the proof of Lemma \ref{1stderiv},
\[f_s''(t)={1\over 2 m_s}\frac{(F_s'(t))^2-F_s''(t)}{F_s^2(t)}\]
and
\begin{eqnarray*}
F_s'(t)&=& -\frac{1}{n!}\int_{K^n} UV\exp(-tU)d\mu,\\
F_s''(t)&=&\frac{1}{n!}\int_{K^n}U^2V\exp(-tU)d\mu.
\end{eqnarray*}
We must show that $(F_s'(t))^2-F_s''(t)\ge0.$ Now, for a fixed $t,$ we may
mulitply $V$ by a constant so that
\[\int_{K^n} V\exp(-tU)d\mu=1.\]
Let $d\gamma :=V\exp(-tU)d\mu.$ Then by the above formulas for $F_s'$ and $F_s'',$ we must show
that
\[\int_{K^n}U^2d\gamma \ge \left(\int_{K^n}U\,d\gamma\right)^2,\]
but this is a simple consequence of the Cauchy-Schwartz inequality. \eop

\newsection{The Limit of Optimal Measures (Designs)}

In this section we state the main theorem. Let $K\subset\C^d$ be  compact with admissible weight function $w:=e^{-\phi}$. 

The class of psh functions of at most logarithmic growth at infinity is denoted by
\[ {\cal L}:=\{u\,:\, u\,\,\hbox{is psh and}\,\,u(z)\le \log^+|z|+C\}.\]

Of special importance is the weighted {\it pluricomplex Green's function} (also known as the weighted
extremal function),
\begin{equation}\label{V}
V_{K,\phi}(z):=
\sup\,\{u(z)\,:\, u\in{\cal L},\,\,u\le \phi\,\,\hbox{on}\,\,K\}.
\end{equation}
The function $V_{K,\phi}^*(z)$ denotes the usc regularization of $V_{K,\phi}.$

Associated to the extremal function is the so-called {\it weighted equilibrium measure},
\begin{equation}\label{wem}
\mu_{K,\phi}:={1\over (2\pi)^d}(dd^cV_{K,\phi}^*)^d.
\end{equation}
Here $(dd^cv)^d$ refers to the non-linear complex Monge-Ampere operator (applied to $v$); it reduces to
(a multiple of) the Laplacian in the dimension $d=1$ case.

That $\mu_{K,\phi}$ exists and is a probability measure
is one of the main results of Pluripotential Theory; we again refer the reader to [K] or [ST] for the details.
We remark, that in one variable, for $K=[-1,1]\subset\C,$ and the unweighted case, i.e., $w=1$ and $\phi=0,$ then
\[ \mu_{K,\phi}={1 \over \pi}{1\over\sqrt{1-x^2}}dx.\]

If $f_1,f_2,\cdots,f_n\in K$ are weighted Fekete points of degree $s$ for $K,$
i.e., points in $K$ for which 
\[|VDM(z_1,\cdots,z_n)|w^s(z_1)w^s(z_2)\cdots w^s(z_n)\]
is maximal, then we may define a discrete probability measure 
\begin{equation}\label{WFekete}
\nu_s={1\over n}\sum_{k=1}^n \delta_{f_k}.
\end{equation}
Berman and Boucksom [BB2] have recently shown that these discrete
probability measures  (\ref{WFekete}) tend weak$-*$ to $\mu_{K,\phi}.$ This is based on a remarkable sequence of papers (see [Be], [BB1], [BB2], [BN]). Indeed, the argument in [BB2] shows  that if for each $s$, we take points $z_1^{(s)},z_2^{(s)},\cdots,z_n^{(s)}\in K$ for which 
\begin{equation}\label{wam}
 \lim_{n\to \infty}\bigl[|VDM(z_1^{(n)},\cdots,z_N^{(n)})|w(z_1^{(s)})^sw(z_2^{(s)})^s\cdots w(z_N^{(s)})^s\bigr]^{1/m_s}=\delta^w(K)
\end{equation}
({\it asymptotically} weighted Fekete points), then the discrete measures 
\[ \nu_n={1\over n}\sum_{k=1}^N \delta_{z_k^{(n)}}\]
converge weak$-*$ to $\mu_{K,\phi}.$ The main point of [BBLW] was to remark
that their proof may be extended to also give the limit of optimal measures (designs). 

\medskip\noindent {\bf Main Theorem.} {\it 
Suppose that $K\subset \C^d$ is compact and that $w$ is an admissible weight function. We  set
$\phi:=-\log(w).$ Suppose further that $\mu_s$ is an optimal measure of degree $s$ for $K$ and $w.$
Then
\[\lim_{s\to\infty}\mu_s=\mu_{K,\phi}\]
where the limit is in the weak$-*$ sense.
}
\medskip

\noindent {\bf Proof.} This is the main result of [BBLW]; its proof is given there. \eop

\newsection{Examples of Equilibrium Measures}

It turns out (see [BT]) that for several important special design spaces $K,$ the 
unweighted ($w=1$) equilibrium measure can be calculated:

\begin{itemize}
\item For $K=[-a,a]^d\subset\R^d,$ a cube, the equilibrium measure is
\[\mu_K=C_d\prod_{i=1}^d\frac{1}{\sqrt{a^2-x_i^2}}dx\]
where $dx$ is Lebesgue measure on $\R^d$ and $C_d$ is the constant,
that depends only on the dimension $d,$ that makes this a probability measure.

\item For $K=\{x\in\R^d\,:\,|x|\le a\},$ the ball of radius $a,$ the equilibrium measure is
\[ \mu_K =C_d a^{-(d-1)}\frac{1}{\sqrt{a^2-|x|^2}}dx\]
where $dx$ is Lebesgue measure on $\R^d$ and $C_d$ is the constant,
that depends only on the dimension $d,$ that makes this a probability measure.

\item For $K=\{x\in\R^d\,:\,x_i\ge 0,\,\, \sum_{i=1}^dx_i\le a\},$ the simplex of ``radius'' $a,$
the equilibrium measure is
\[\mu_K=C_d a^{-(d-1)/2} \frac{1}{\sqrt{(a-\sum_{i=1}^dx_i)\prod_{i=1}^dx_i}}dx\]
where $dx$ is Lebesgue measure on $\R^d$ and $C_d$ is the constant,
that depends only on the dimension $d,$ that makes this a probability measure.
\end{itemize}
We also offer a weighted example. Take
\[K=\{z \in \C^d\, :\, |z|\le 1\}\]
with $\phi(z)=|z|^2.$ Then it can be verified that the
extremal function is 
\[V_{K,\phi}(z)=\left\{ \begin{array}{cl}
\phi(z)=|z|^2&\hbox{if}\,\,|z|\le 1/\sqrt{2} \cr
\log(|z|)+1/2-\log(1/\sqrt{2})&\hbox{otherwise}
\end{array}.\right. \]
From this one may readily compute
$\mu_{K,\phi}$ is Lebesgue measure supported on the
ball $\{z \in \C^d\, :\, |z|\le 1/\sqrt{2}\},$ normalized to
be a probability measure.
\bigskip

%%%%%%%%%%%%%%%%References %%%%%%%%%%%%%%%%%%%

\centerline {\bf REFERENCES}
\vskip10pt
\begin{description}
\item {[BT]} Bedford, E., and Taylor, B.~A., {\it The complex equilibrium measure of a symmetric convex set in
$\R^n$}, Trans. AMS, 294, 705--717.

\item{[Be]} Berman, R., {\it Bergman Kernels for Weighted Polynomials and Weighted Equilibrium Measures
of $\C^n$}, preprint.

\item {[BB1]} Berman, R. and Boucksom, S., {\it Capacities and Weighted Volumes of Line Bundles},
preprint.

\item {[BB2]} Berman, R. and Boucksom, S., {\it Equidistribution of Fekete Points on Complex Manifolds},
preprint.

\item{[BB3]}  R. Berman and S. Boucksom, Growth of balls of
holomorphic sections and energy at equilibrium, \emph{Invent. Math.}, \textbf{181} (2010),  no. 2, 337-394.

\item {[BN]} Berman, R. and Nystrom, D.W., {\it Convergence of Bergman Measures for High Powers of a
Line Bundle}, preprint.

\item{[BBN]} R. Berman, S. Boucksom and D. W. Nystrom, Convergence towards equilibrium on complex manifolds, to appear in \emph{Acta Math.}

%\item {[Bh]} Bhatia, R., {\bf Matrix Analysis}, GTM 169, Springer, New York, 1997.

\item{[BBCL]} Bloom, T., Bos, L., Christensen, C. and Levenberg, N.,  {\it Polynomial interpolation of holomorphic functions  in $\C$ and $\C^n$}, Rocky Mtn. J. Math 22 (1992),  441--470.

\item{[BBLW]} Bloom, T., Bos, L., Levenberg, N. and Waldron, S., {\it On the Convergence of Optimal Measures}, to appear in
Constr. Approximation.

\item{[BL]} Bloom, T. and Levenberg, N., {\it Transfinite diameter notions in $\C^n$ and integrals of Vandermonde
determinants}, preprint.

\item {[Bo]} Bos, L., {\it Some Remarks on the Fej\'er Problem for Lagrange Interpolation in Several Variables},
J. Approx. Theory, Vol. 60, No. 2 (1990), 133 -- 140.

\item {[D]} Deift, P., {\bf Orthogonal polynomials and random matrices: a Riemann-Hilbert approach}, AMS, 1998.

\item{[DS]} Dette, H. and Studden, W.J., {\bf The Theory of Canonical Moments with Applications in Statistics,
Probability and Analysis}, Wiley Interscience, New York, 1997.

%\item {[F]} Fej\'er, L., {\it Bestimmung dergenigen Abszissen eines Intervalles f\"ur welche die
%Quadratsumme der Grundfunktionen der Lagrangeschen Interpolation im Intervalle eing moglichst kleines Maximum besitzt},
%Ann. Scuoal Norm. Sup. Pisa (2) {\bf 1} (1932), 263 -- 276.

\item{[KS]} Karlin, S. and Studden, W.J.,  {\bf Tchebycheff Systems: With Applications in Analysis and Statistics}, Wiley Interscience, New York, 1966.

\item{[KW]} Kiefer, J. and Wolfowitz, J., {\it The equivalence of two extremum problems}, Canad. J. Math. {\bf 12}
(1960), 363 -- 366.

\item{[K]} Klimek, M., {\bf Pluripotential Theory}, Oxford Univ. Press, 1991.

%\item{[R]} Rumely, R., {\it A Robin Foumula for the Fekete-Leja Transfinite Diameter}, Math. Ann. {\bf 337} no. 4 (2007), 729 -- 738.

\item{[ST]} Saff, E. and Totik, V., {\bf Logarithmic Potentials with External Fields}, Springer, 1997.

\item{[Z]} Zaharjuta, V.~P.,  {\it Transfinite diameter, Chebyshev constants, and capacity for compacta in $\C^n$}, Math. USSR Sbornik, {\bf 25} (1975), no. 3, 350 -- 364.

\end{description}

\end{document}